\documentclass[
    reprint,
    superscriptaddress,
    amsmath, amssymb,
    aps,
]{revtex4-2}

\usepackage{natbib}
\usepackage{graphicx}
\usepackage{dcolumn}
\usepackage{bm}

\usepackage[T1]{fontenc}
\usepackage[english]{babel}
\usepackage{physics}

\usepackage[position=top, font=normal, labelfont=bf]{subcaption}

\usepackage{ragged2e}

\newcommand{\mrm}[1]{\mathrm{#1}}
\newcommand{\be}{\begin{equation}}
\newcommand{\ee}{\end{equation}}
\usepackage{xcolor}

\begin{document}
\title{Spatial and temporal cluster tomography of active matter}
\author{Leone~V.~Luzzatto}
\affiliation{Department of Physics and Astronomy, Northwestern University, Evanston, IL 60208}
\affiliation{NSF-Simons National Institute for Theory and Mathematics in Biology, Chicago, IL 60611}
\author{Mathias~Casiulis}
\affiliation{Center for Soft Matter Research, New York University, NY 10003}
\affiliation{Simons Center for Computational Physical Chemistry, New York University, NY 10003}
\author{Stefano~Martiniani}
\affiliation{Center for Soft Matter Research, New York University, NY 10003}
\affiliation{Simons Center for Computational Physical Chemistry, New York University, NY 10003}
\affiliation{Courant Institute of Mathematical Sciences, New York University, NY 10003}
\affiliation{Center for Neural Science, New York University, NY 10003}
\author{Istv\'{a}n~A.~Kov\'{a}cs}
\affiliation{Department of Physics and Astronomy, Northwestern University, Evanston, IL 60208}
\affiliation{NSF-Simons National Institute for Theory and Mathematics in Biology, Chicago, IL 60611}
\affiliation{Northwestern Institute on Complex Systems, Northwestern University, Evanston, IL 60208}
\affiliation{Department of Engineering Sciences and Applied Mathematics, Northwestern University, Evanston, IL 60208}

\begin{abstract}
Critical phase transitions have proven to be a powerful concept to capture the phenomenology of many systems, including deeply non-equilibrium ones like living systems.
The study of these phase transitions has overwhelmingly relied on two-point correlation functions.
In this Letter, we show that cluster tomography---the study of one-dimensional cross-sections of the clusters that emerge near a phase transition---is an alternative higher-order tool that efficiently locates and characterizes phase transitions in active systems.
First, using motility-induced phase separation as a paradigmatic example, we show how complex geometric features of clusters, captured by \emph{spatial} cluster tomography, can be used to measure critical exponents in active systems without explicitly introducing system-specific order parameters.
Second, we introduce \emph{temporal} cluster tomography, an analogous cluster-based measurement that characterizes the dynamical behavior of active systems.
We show that cluster dynamics can be captured by a generalization of burstiness analysis in complex temporal networks.
Both spatial and temporal cluster tomography are easy to implement yet powerful approaches to study non-equilibrium systems, making them useful additions to the standard toolbox of statistical physics.
\end{abstract}
\maketitle


\renewcommand{\figurename}{FIG.}

Active systems are driven out of equilibrium by a sustained conversion of energy to work at the microscopic scale.
This leads to emergent collective phenomena with symmetries, conservation laws, and material properties radically different from those found in equilibrium~\cite{Marchetti2013, ginelli2016, ramaswamy2017,Fruchart2021,Baconnier2022}.
Yet, like equilibrium systems~\cite{book:goldenfeld1992, book:kardar2007}, active systems may display critical phase transitions associated to a diverging correlation length,~\textit{e.g.}~isotropic-polar transitions~\cite{vicsek1995}, or motility-induced phase separations (MIPS)~\cite{Cates2015}.
This divergence is often studied through the properties of geometric domains, or \emph{clusters}, \textit{e.g.} of aligned spins, or of high-density regions.
Cluster-based analyses have found success in equilibrium classical~\cite{FK1972,Baxter1982} and quantum~\cite{love2025} systems, as well as non-equilibrium physical~\cite{bernard2006, henkes2016, zhang2019} and biological~\cite{be'er2020, ansell2024} systems.
Standard properties of clusters include their size distribution, fractal dimension, and boundary length---all governed by two-point correlation functions~\cite{stauffer1994introduction}. 
In contrast, \emph{cluster tomography}, a measurement inspired by entanglement in disordered quantum systems~\cite{Yu2008,kovacs2012universal}, incorporates deeper information about higher-order correlation functions~\cite{Ansell2023}, making it a powerful tool to characterize phases and the transitions between them.

%
%
\begin{figure}[ht!]
\begin{center}
    \includegraphics[width=\linewidth]{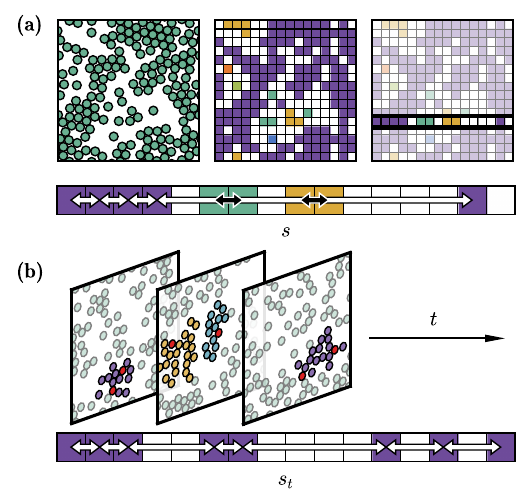}
\end{center}

\begin{subcaptiongroup}
\phantomcaption\label{fig:explanation.spatial}
\phantomcaption\label{fig:explanation.temporal}
\end{subcaptiongroup}
\vspace{-10pt}

\caption{
\label{fig:explanation}
\justifying
\textbf{Cluster tomography.} \textbf{(a)} Spatial cluster tomography. Given a cluster-forming system, we consider one-dimensional cross-sections and tally the distance $s$ between successive observations of each cluster, to detect and characterize phase transitions. 
\textbf{(b)} Temporal cluster tomography. Given a pair of particles (red), we measure the time intervals $s_t$ between successive observations of them belonging to the same cluster (first and third snapshots) to study the dynamics of the system.
}
\end{figure}
%
%

In this Letter, we apply cluster tomography to Active Brownian Particles (ABPs)~\cite{Fily2012,CristinaMarchetti2016} undergoing MIPS~\cite{Cates2015}. 
We study two regions of the phase diagram, demonstrating how cluster tomography can be used to detect and characterize both first- and second-order phase transitions.
Furthermore, we introduce an analogous notion of \emph{temporal} cluster tomography, and a generalization of ``burstiness analysis'' in complex networks~\cite{goh2008burstiness, karsaibook} that can provide insight into the dynamical behavior in different phases.
We show that spatial and temporal cluster tomography constitute  powerful, easy-to-implement tools to study the static and dynamical properties of active systems from a collection of snapshots.

\emph{Spatial cluster tomography}---The procedure for spatial cluster tomography is sketched in Fig.~\ref{fig:explanation.spatial}.
In short, given a definition for clusters (here, particle proximity), we consider one-dimensional cross-sections of the clustered system, which can be represented as sequences of cluster labels. 
From these sequences, we measure the distance $s$ between successive observations of the \textit{same} cluster, which we refer to as a spatial \emph{gap}.
We then define the gap-size statistics $g(s)$ as the number of gaps of size $s$ in one such sequence.
By definition, a spatial gap of size $s$ carries information about the cluster membership of $s$ points along a line, so $g(s)$ is a higher-order measurement than the two-point connectivity.
This procedure is reminiscent of the ``chord-length'' approach to the study of binary random media~\cite{Levitz1992, Torquato1993, zhang2025}, in which a binary label identifies the phase at each site but individual clusters are not distinguished.
Identifying each cluster is however crucial to characterizing the critical behavior.
In equilibrium systems~\cite{kovacs2012corner,kovacs2014corner,kovacs2014excess}, the gap-size statistics was found to decay exponentially in phases with a finite spatial correlation length, while it is a power-law at criticality: 
\be
\label{gapsize_statistics}
\hat{g}(s)\equiv\frac{g(s)}{L} = b\,s^{-\zeta}\;,
\ee
where $L$ is the linear size of the system.
In two dimensions, most critical points studied so far are consistent with a hyper-universal exponent value of $\zeta=2$, which can often be attributed to a conformally invariant scaling limit (an exception being the disordered quantum Ising model~\cite{love2025}).
Crucially, the prefactor $b$ is also universal, with its value characterizing the universality class of the transition~\cite{kovacs2014corner}.
We refer to $b$ as the \emph{lacunarity}, as it characterizes the tendency of clusters to wrap around empty regions leading to intricate cluster shapes with holes and/or concave boundaries.

As discussed in detail in Appendix~\ref{appendix.theory}, a non-zero lacunarity leads to the special case of a logarithmic ``corner contribution'' $\mathcal{C}=b\ln\ell$ in the number of clusters intersecting a line of length $\ell$ at criticality~\cite{Yu2008,kovacs2013infinitely}.
The corner contribution has been studied in a number of equilibrium classical~\cite{kovacs2012corner,kovacs2014corner,kovacs2014excess} and quantum~\cite{kovacs2012universal,bueno2015,helmes2016,love2025} systems in the vicinity of a critical point for the more general case of contour lines with various corner angles---\textit{e.g.} squares and triangles (a line segment corresponds to two corners with $2\pi$ angles).
Here, by means of the gap-size statistics, we demonstrate that corner contributions also yield meaningful insight in active systems.

\emph{Temporal cluster tomography}---We now propose a concept analogous to spatial cluster tomography in the time domain.
Instead of the geometric distance between consecutive observations of a cluster, we consider time intervals between consecutive ``cluster events.''
Na\"ively, one could imagine tracking the cluster label at a fixed location, or the cluster membership of a fixed particle. 
However, clusters form, merge, split, and dismantle over time, making cluster identity ambiguous across snapshots.
To overcome this difficulty, we instead define relevant events as two particles belonging to the same cluster.
A temporal gap $s_t$ is then given by the time interval between such events, as sketched in Fig.~\ref{fig:explanation.temporal}.
We define the temporal gap-size statistics $g_t(s_t)$ by accumulating $s_t$ values across particle pairs over a given time window.

This notion of temporal gaps is reminiscent of measurements in complex temporal networks, \textit{e.g.} the study of time intervals between communication events between pairs of agents, such as two people being on the same phone call.
Thus, as a candidate for an informative summary statistics, we propose to use a tool from complex temporal networks, known as ``burstiness'' analysis~\cite{goh2008burstiness, karsaibook}.
In systems with no dynamical complexity (\textit{i.e.}~no memory) inter-event times are expected to follow an exponential distribution~\cite{goh2008burstiness}, deviations from which are conveniently quantified by the \emph{burstiness parameter},
\be
    B = \frac{ \sigma_t - \expval{s_t} }{ \sigma_t + \expval{s_t} }\;,
\ee
where $\expval{s_t}$ and $\sigma_t$ stand for the mean and standard deviation of the inter-event time distribution, respectively.
$B > 0$ describes bursty dynamics, often characterized by avalanche-like events with a broad inter-event time distribution.
$B<0$, on the other hand, describes more regular dynamics, with temporal gaps narrowly distributed around a characteristic inter-event time.
Like the lacunarity $b$, the burstiness parameter $B$ is a simple summary statistics that helps detect and characterize transitions between dynamical regimes.

\emph{The model}---To showcase the potential of cluster tomography in the context of active systems, we focus on MIPS, a paradigmatic example of a collective phenomenon unique to active systems~\cite{Cates2015}, by which self-propelled particles with purely repulsive interactions can spontaneously separate into dense and dilute fluid phases~\cite{tailleur2008,Fily2012,Redner2013}.
The mechanism of MIPS is now well understood---in a nutshell, phase separation is driven by a feedback loop in which collisions cause particles to move more slowly through high-density regions and, in turn, active particles tend to accumulate in regions where they move more slowly~\cite{Bialke2013,Cates2015,CristinaMarchetti2016}.
While MIPS has been studied extensively, its critical behavior is still debated.
Some numerical investigations found results consistent with a critical point in the Ising universality class~\cite{partridge2019, maggi2021}, while others, using slightly different models, report significant deviations from some Ising critical exponents~\cite{siebert2018, dittrich2021}.
Renormalization group studies paint a nuanced picture, hinting at an Ising-like fixed point that is nonetheless significantly altered by the non-equilibrium nature of the model~\cite{caballero2018, speck2022}.
Finally, it was shown that MIPS arises from effective many-body interactions~\cite{Farage2015,ArnoulxDePirey2025}, so that going beyond pair correlations may be necessary to fully characterize its criticality.
MIPS is thus an ideal test case for cluster tomography: its mechanism and domain are well understood, yet using a new higher-order tool may resolve controversies on its critical behavior.

Concretely, we consider two-dimensional ABPs, \textit{i.e.} polar particles with diameter $a$ and self-propulsion speed $v_0$, interacting through finite-range harmonic repulsion and subject to rotational noise.
A full description of our choice of ABP dynamics can be found in Appendix~\ref{appendix.ABP}.
We simulate $N$ particles in a square box with periodic boundary conditions and side-length $L$.
Two dimensionless parameters determine together whether MIPS is observed~\cite{Fily2014,Nie2020}: the volume fraction occupied by the disks, $\phi=N\pi a^2/L^2$, and the ratio of the self-propulsion and rotational diffusion time-scales, often called rotational P\'eclet number, $\mathrm{Pe}_r = v_0 / (aD_r)$, where $D_r$ is the rotational diffusion coefficient.
In the $(\phi,\mathrm{Pe}_r)$ plane, MIPS is observed in a U-shaped region at large angular P\'eclet numbers and intermediate volume fractions~\cite{Nie2020}. See Supplementary Material (SM) for a full phase diagram~\cite{[{See Supplemental Material at }][{ for a complete description of numerical methods, a full phase diagram of our choice of ABP dynamics, and additional notes on the implementation of cluster tomography.}]supp}.

%
%
\begin{figure}[t]
\begin{center}
    \includegraphics[width=\linewidth]{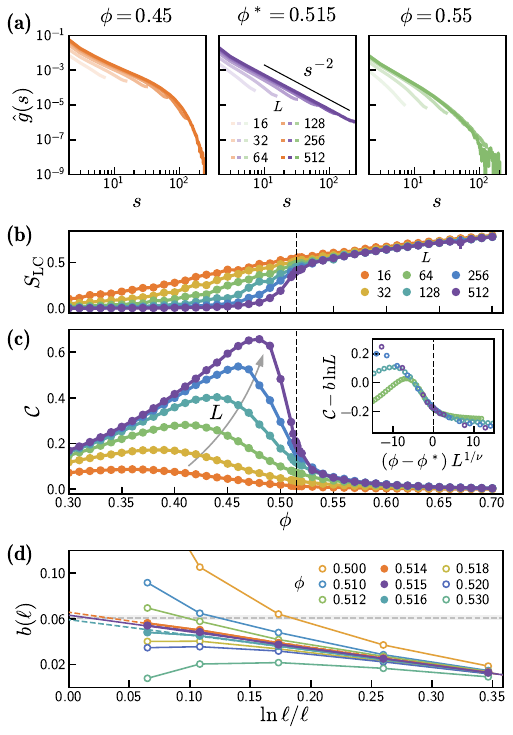}
\end{center}

\begin{subcaptiongroup}
\phantomcaption\label{fig:spatial.gapsize}
\phantomcaption\label{fig:spatial.largest}
\phantomcaption\label{fig:spatial.corner}
\phantomcaption\label{fig:spatial.b}
\end{subcaptiongroup}
\vspace{-15pt}

\caption{
\label{fig:spatial}
\justifying
\textbf{Spatial cluster tomography.} 
\textbf{(a)} Gap-size statistics at $\mathrm{Pe}_r=10$ and volume fractions below, close to, and above the critical region. 
\textbf{(b)} Fraction of the available volume occupied by the largest cluster. 
\textbf{(c)} Corner contribution $\mathcal{C}$, excluding the largest cluster. The dashed line marks $\phi^* = 0.515$.
(Inset) Data collapse of the corner contribution using the Ising values $b=0.0608$, $\nu=1$. $L \leq 32$ data are omitted for clarity. 
\textbf{(d)} Finite-size estimates of the lacunarity using $\ell=L/8$. An infinite-size extrapolation in $\ln\ell/\ell$ shows consistency with the Ising lacunarity (dashed horizontal line).
Where no error bars are visible, the uncertainty is smaller than the marker.
}
\end{figure}
%
%

\emph{Spatial tomography at criticality}---First, we focus on simulations with fixed $\mathrm{Pe}_r = 10$, close to the low-$\mathrm{Pe}_r$ boundary of the MIPS region of the phase diagram~\cite{Nie2020}, and use the volume fraction as the control parameter.
We consider linear system sizes between $L=16$ and $512$ and volume fractions between $\phi=0.3$ and $0.7$.
For each $(L,\phi)$ pair, we measure the gap-size statistics using $10^2$ to $10^5$ snapshots. 
In each snapshot, we identify clusters via particle proximity: particle positions are binned using a grid of square cells with side length $a$, and neighboring occupied cells are taken to be in the same cluster (Fig.~\ref{fig:explanation.spatial}).
Following the standard prescription for percolation processes~\cite{stauffer1994introduction}, we exclude the largest cluster from these measurements (see SM~\cite{supp}).

Our results are shown in Fig.~\ref{fig:spatial}.
Close to $\phi^*=0.515$ the gap-size statistics follows a power-law distribution characterized by an exponent close to $\zeta=2$, the hyper-universal value often observed in close proximity to a critical point (Fig.~\ref{fig:spatial.gapsize}).
For $\phi>\phi^*$, the gap-size statistics is well described by a power-law with an exponential cutoff, $\hat{g}(s)\sim s^{-\zeta}e^{-s/s_0}$, that eventually becomes a pure exponential decay far from $\phi^*$.
At $\phi^*$, we also observe a sharp increase in the size of the largest cluster, consistent with a transition between a sparse and a dense phase (Fig.~\ref{fig:spatial.largest}). 
The largest cluster data exhibits large fluctuations despite the large number of samples studied, illustrating the challenges that arise with standard approaches.

The corresponding corner contribution $\mathcal{C}$, calculated from the gap-size statistics, is shown in Fig.~\ref{fig:spatial.corner}.
We observe a peak close to $\phi^*=0.515$ that grows with increasing system size.
This is consistent with the presence of a critical point: in equilibrium systems characterized by $\zeta=2$, $\mathcal{C}$ diverges logarithmically with $L$ at criticality, while it is expected to be bounded outside of the critical region~\cite{Yu2008,kovacs2012corner,Ansell2023,kovacs2012universal}.

We now study the lacunarity $b$ to determine the universality class of the critical point.
We extract finite-size estimates of $b$ from the gap-size statistics using (Appendix~\ref{appendix.theory}) 
\be\label{b_finite}
    b(\ell) = \frac{1}{\ln\qty(1+\frac{1}{\ell})} \qty[ 2\sum_{s=\ell+1}^{2\ell} \hat{g}(s) + \hat{g}(2\ell+1) ]\;.
\ee
Figure~\ref{fig:spatial.b} shows results for $\ell=L/8$ (this exact value is irrelevant as long as $\ell/L=\mathcal{O}(1)$, and the analysis is robust over a wide range $L/16 \leq \ell \leq L/5$~\cite{supp}).
Extrapolating to the infinite-size limit, $\ln\ell/\ell\to0$, we estimate the lacunarity $b$ for the MIPS critical point to be $b_\mathrm{ABP} = 0.0623(27)$.
The only known model with a lacunarity compatible with this value is the Ising model, for which $b_\mrm{\,Ising}=0.0608(13)$~\cite{kovacs2014corner}.
To further support the Ising universality class hypothesis, we study the correlation length exponent $\nu$.
We find a good data collapse of the corner contribution using the Ising value $\nu=1$~\cite{book:goldenfeld1992}
(Fig.~\ref{fig:spatial.corner}, inset).
Additionally, fitting the gap-size statistics above $\phi^*$ with $\hat{g}(s)\sim s^{-2}e^{-s/s_0}$, we find that $s_0$ scales with the volume fraction as $s_0(\phi)\sim(\phi-\phi^*)^{-\nu}$, with $\nu=1.08(5)$~\cite{supp}.

\emph{Temporal tomography at criticality}---We now turn our attention to temporal tomography.
We use $10^3$--$10^4$ snapshots in steady state, taken at regular time intervals $\Delta t = a/v_0$ apart.
Since considering all ${N}\choose{2}$ pairs of $N\propto L^2$ particles would become rapidly infeasible with increasing $L$, we select a random sample of $M \propto \sqrt{N}$ particles and consider all possible pairs between them, as this choice already provides sufficient statistics.

%
%
\begin{figure}[t]
\begin{center}
    \includegraphics[width=\linewidth]{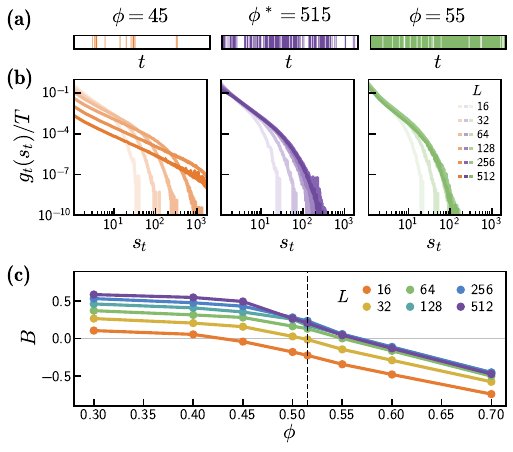}
\end{center}

\begin{subcaptiongroup}
\phantomcaption\label{fig:time.series}
\phantomcaption\label{fig:time.gapsize}
\phantomcaption\label{fig:time.burstiness}
\phantomcaption\label{fig:time.infinite}
\end{subcaptiongroup}
\vspace{-15pt}

\caption{
\label{fig:time}
\justifying
\textbf{Temporal cluster tomography.} \textbf{(a)} Representative time series for a particle pair at volume fractions below, at, and above the critical region over 200 time steps ($L=512$). Vertical lines mark snapshots where the pair is in the same cluster. 
\textbf{(b)} Temporal gap-size statistics at the same volume fractions as (a), normalized by the number of snapshots $T$.
\textbf{(c)} Burstiness parameter calculated from $g_t(s_t)$.
To avoid artefacts due to the periodic boundaries, only temporal gaps $s_t<L$ are used in the calculation.
The dashed line marks $\phi^*=0.515$.
Where no error bars are visible, the uncertainty is smaller than the marker.
}
\end{figure}
%
%

The temporal gap-size statistics $g_t(s_t)$, shown in Fig.~\ref{fig:time.gapsize}, displays a clear qualitative change close to $\phi^*=0.515$.
At low volume fractions,  $g_t(s_t)$ is found to be heavy-tailed, indicating that large temporal gaps are common, especially as the system size increases.
Above $\phi^*$, only short gaps are present, and increasing the system size does not alter the observed statistics (Fig.~\ref{fig:time.gapsize}).
This qualitative change in the dynamics translates into a shift in the burstiness parameter $B$ from a positive to a negative value near $\phi^*$ (Fig.~\ref{fig:time.burstiness}).
At low volume fractions, we find a large positive $B$ that increases with the system size and remains approximately constant for varying $\phi$.
In contrast, above $\phi^*$, $B$ exhibits a strong dependence on $\phi$ but not on $L$ and quickly becomes negative at higher volume fractions.

Our results can be interpreted in light of the correlation length of the system.
Below $\phi^*$, cluster size is bounded by the finite correlation length $\xi$.
As long as the distance between two particles is larger than $\xi$, they are unlikely to be found in the same cluster.
However, if the distance between them becomes comparable to $\xi$, they will often switch between belonging to the same cluster then to different ones, as clusters continuously form, merge, and break apart.
We observe this as sequences of short temporal gaps separated by long temporal gaps, typical of bursty ($B > 0$) dynamics (Fig.~\ref{fig:time.series}).
On the other hand, when $\phi\gg\phi^*$, the system essentially contains a single cluster with empty pockets.
When a particle leaves the cluster, the typical gap time is set by the typical size of empty regions in the system and the self-propulsion speed.
As a result, the frequency of temporal gaps drops sharply after $s_t=1$, leading to $B<0$.

\begin{figure}[t]
\begin{center}
\includegraphics[width=\linewidth]{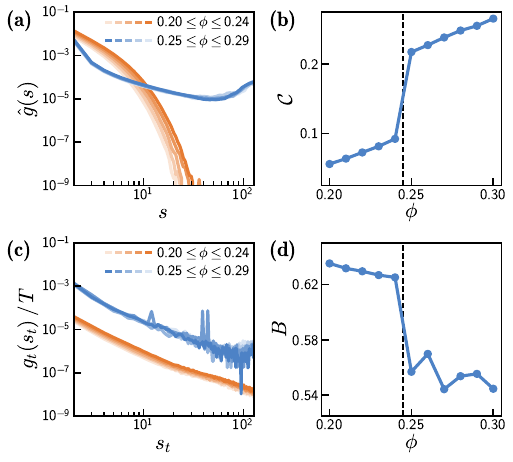}
\end{center}

\begin{subcaptiongroup}
\phantomcaption\label{fig:first.spatial_gapsize}
\phantomcaption\label{fig:first.corner}
\phantomcaption\label{fig:first.dynamical_gapsize}
\phantomcaption\label{fig:first.burstiness}
\end{subcaptiongroup}
\vspace{-15pt}

\caption{
\label{fig:first}
\justifying
\textbf{First-order transition.} Gap-size statistics~(a), corner contribution~(b), temporal gap-size statistics~(c), and burstiness parameter~(d) across the MIPS boundary at $\mathrm{Pe}_r=100$, $L=256$.
}
\end{figure}

\emph{First-order transition}---Finally, we demonstrate that cluster tomography is not limited to the study of continuous phase-transitions. 
To do so, we carry out a similar analysis on simulations with $\mathrm{Pe}_r=100$, much larger than the value at the MIPS critical point at $\mathrm{Pe}_r\approx10$~\cite{Nie2020,supp}.
As we vary the volume fraction between $\phi=0.2$ and $0.3$, we see a sudden qualitative change in the shape of the gap-size statistics and a corresponding jump in the corner contribution (Figs.~\ref{fig:first.spatial_gapsize} and~\ref{fig:first.corner}), consistent with the presence of a first-order phase transition~\footnote{Note that the largest cluster was included in this analysis, as its exclusion was only helpful to mitigating finite-size effects close to the critical point.}.
We observe a similar sudden change in the shape of the temporal gap-size statistics, and a corresponding jump in the burstiness parameter, marking the transition to a different dynamical regime (Figs.~\ref{fig:first.dynamical_gapsize} and~\ref{fig:first.burstiness}).
We note that, in the phase-separated regime, $B$ is sensitive to changes in the cluster structure---\textit{e.g.} the largest cluster temporarily splitting in two---leading to a larger variability, not entirely accounted for by our estimate of the uncertainty.

%
%
%
%

\emph{Discussion}---Spatial cluster tomography is a powerful tool to measure universal higher-order geometric correlations in active systems.
Studying the gap-size statistics, the corner contribution, as well as the lacunarity, we located and characterized first and second-order phase transitions in ABPs.
We showed that the near-critical behavior is compatible with a critical point in the Ising universality class, matching both the Ising lacunarity $b$ and correlation length exponent $\nu$.
Our work constitutes the first measurement of the corner contribution in an out-of-equilibrium system, and shows that corner effects in active systems are a promising subject for future research.
In particular, whether the angle-dependence of corner effects in active systems follows the equilibrium theory~\cite{cardy1988,estienne2022} should be investigated to probe the emergent symmetries of these systems.

We also introduced temporal cluster tomography to probe dynamical properties across phases.
By analogy with complex networks, we introduced the burstiness parameter $B$~\cite{goh2008burstiness,karsaibook}, an informative and easily accessible summary statistics that can be used to characterize active dynamics.
We reported a heavy-tailed, bursty ($B>0$) distribution of temporal gaps in the sparse (low--volume fraction) phase, and a more regular ($B<0$) dynamics in the high-volume fraction phase.
While we observed $B\approx0$ at intermediate volume fractions, this value does not appear to correspond to a physically significant point in parameter space, which highlights a potential limitation of the burstiness parameter as a summary statistics for temporal cluster tomography.
Future work could identify alternative metrics to capture the rich information contained in the temporal gap-size statistics.

Cluster tomography is applicable to any cluster-forming system in physics, biology and beyond---and thus joins an expanding family of promising model-agnostic characterization methods~\cite{Martiniani2019,Martiniani2020,Cavagna2021,Ro2022}.
Especially relevant are systems expected or conjectured to be conformally invariant, such as turbulent flow~\cite{bernard2006}, rigidity transitions~\cite{Javerzat2023}, and living matter~\cite{Andersen2025}, as such symmetries lead to clear theoretical expectations for the lacunarity and corner contribution~\cite{kovacs2012corner,bueno2015}.

\begin{acknowledgments}
This work was supported by the National Science Foundation under Grant No.~PHY-2310706 of the QIS program in the Division of Physics, and by the Baker Faculty Grant of the Weinberg College of Arts and Sciences, Northwestern University.
This research was supported in part through the computational resources and staff contributions provided for the Quest high performance computing facility at Northwestern University, which is jointly supported by the Office of the Provost, the Office for Research, and Northwestern University Information Technology.
We thank Monica Olvera de la Cruz, Daniel Matoz-Fernandez, Michelle Driscoll, Brendan Blackwell, Helen S.~Ansell, Sean P. E. Dougherty, and Xiangyi Meng for useful discussions.
\end{acknowledgments}

\bibliography{references,supp}

\appendix
\setcounter{figure}{0}
\setcounter{equation}{0}
\renewcommand{\figurename}{FIG.}
\renewcommand{\thefigure}{A\arabic{figure}}
\renewcommand{\thetable}{A\arabic{table}}
\renewcommand{\theequation}{A\arabic{equation}}

\section{Gap sizes and the corner contribution}\label{appendix.theory}

In the main text, we focus on measurements of the gap size statistics.
An alternative quantity that can be studied in clustered systems is the average number of distinct clusters $N(\ell)$ encountered along a linear path of length $\ell$~\cite{Yu2008,kovacs2012corner}.
If the correlation length is finite, up to leading order, $N(\ell)$ scales linearly with the length of the path.
Close to a critical point, however, an additional, non-linear correction $\mathcal{C}(\ell)$ is often observed, so that $N(\ell) = a\ell + \mathcal{C}(\ell)$~\cite{Yu2008,kovacs2012corner,kovacs2014corner,kovacs2014excess,Ansell2023}.
This nonlinear correction is known as the \emph{corner contribution}, as in general it originates from sharp corners in the contour along which clusters are counted---the endpoints in the case of a line segment.
The leading order of the corner contribution is often logarithmic,
\begin{equation}
\label{corner}
\mathcal{C}(\ell) = b\ln{\ell} + \mathrm{const.}\;,
\end{equation}
and when this is the case, the constant $b$ was found to be universal.
Throughout the main text, we have used the notational shorthand $\mathcal{C}(L/2)\equiv\mathcal{C}$ when discussing the corner contribution, where $L$ is the linear size of the system.

The corner contribution and gap size statistics are different aspects of the same underlying physics.
In systems with periodic boundary conditions, given a path length $\ell=L/2$, they are related exactly via
\cite{hoyos2007,Yu2008,kovacs2012universal,zou2022,Ansell2023}
\begin{equation}
\label{double_sum}
\mathcal{C}(L/2)=\frac{1}{2}\sum_{r=1}^{L/2}\sum_{s=r}^{L - r}\hat{g}(s)\;.    
\end{equation}
This relationship is analogous to the one between the chord-length distribution and the lineal-path function~\cite{lu1992,Torquato1993} that has been studied in the context of two-phase random media.
It follows from Eq.~\ref{double_sum} that a logarithmic corner contribution corresponds to a power-law gap-size statistics with the hyper-universal exponent $\zeta=2$ that is often observed at critical points, and that  the universal constant $b$ in Eq.~\ref{corner} is the same universal lacunarity that governs the gap-size statistics (Eq.~\ref{gapsize_statistics}).

If the leading order of the corner contribution is logarithmic, it can be computed from the gap-size statistics for arbitrary $\ell$ as
\begin{equation}
\label{double_sum_ell}
\mathcal{C}(\ell)=\sum_{r=1}^{\ell}\sum_{s=r}^{2\ell - r}\hat{g}(s)\;.
\end{equation}
Our formula to calculate finite-size estimates of $b$ (Eq.~\ref{b_finite}) follows directly from Eqs.~\ref{corner} and~\ref{double_sum_ell} using
\begin{equation}
b(\ell) = \pdv{\,\mathcal{C}(\ell)}{\ln\ell} \approx \frac{\mathcal{C}(\ell+1)-\mathcal{C}(\ell)}{\ln(\ell+1)-\ln(\ell)}\;.
\end{equation}

\setcounter{figure}{0}
\setcounter{equation}{0}
\renewcommand{\figurename}{FIG.}
\renewcommand{\thefigure}{B\arabic{figure}}
\renewcommand{\thetable}{B\arabic{table}}
\renewcommand{\theequation}{B\arabic{equation}}

\section{Implementation of ABP dynamics}\label{appendix.ABP}

Our model of ABPs consists of polar particles with positions $\vb{r}_i$ and polarities $\vu{n}_i = (\cos\theta_i, \sin\theta_i)$ whose dynamics follow 
\begin{align}
    \partial_t{\vb{r}_i} &= v_0\vu{n}_i(\theta_i) + \mu \sum_{j\neq i} \vb{F}_{ij}(r_{ij}) \;, &
    \partial_t{\theta_i} &= \sqrt{2D_r} \, \xi_i(t) \;, \label{eq:ABPs}
\end{align}
where $v_0$ is the self-propulsion speed, $\mu$ is the mobility coefficient, $D_r$ is the rotational diffusion coefficient, and $\xi_i$ denotes zero-mean, unit-variance Gaussian noise.
We set inter-particle forces to be finite-range harmonic repulsion, $\vb{F}_{ij} = k(a - r_{ij})\vu{r}_{ij}$ if $r_{ij} < a$, and $0$ otherwise, where $k$ is a stiffness constant, $a$ is the diameter of the repulsive disks, $r_{ij} = \abs{\vb{r}_i - \vb{r}_j}$, and $\vu{r}_{ij} = (\vb{r}_i - \vb{r}_j)/r_{ij}$.
We set the softness of the repulsion via $\mu k a/v_0 = 10^2$.
A detailed discussion of our numerical simulations of Eq.~\ref{eq:ABPs} can be found in the SM~\cite{supp}.

\end{document}